\documentclass[lettersize,journal,twoside]{IEEEtran}
\usepackage{amsmath,amsfonts, amssymb}
\usepackage{subfig}
\usepackage{textcomp}
\usepackage{url}
\usepackage{verbatim}
\usepackage{booktabs} 
\usepackage{graphicx}
\usepackage{cite}
\usepackage{amsmath}
\usepackage{graphicx}
\usepackage{xcolor}
\usepackage{ifthen}
\usepackage{comment}
\usepackage{enumitem} 
\usepackage[hidelinks]{hyperref}
\newcommand{\code}[1]{\textsf{\small#1}}
\newboolean{showcomments}
\ifdefined\DRAFT
  \setboolean{showcomments}{true} 
\else
  \setboolean{showcomments}{false}
\fi
\makeatletter
\newcommand{\mynote}[3]{%
  \ifthenelse{\boolean{showcomments}}{%
    {\textbf{\textcolor{#3}{(#1 $\triangleright$ #2)}}}}%
  {%

  }%
}
\makeatother

\newcommand{\fixme}[1]{\mynote{FIXME}{#1}{red}}

\newcommand{\eal}[1]{\mynote{Edward}{#1}{blue}}
\newcommand{\erj}[1]{\mynote{Erling}{#1}{cyan}}
\newcommand{\peter}[1]{\mynote{Peter}{#1}{magenta}}

\usepackage[scaled=1.04]{couriers}
\definecolor{stringColor}{rgb}{0.1,0.5,0.1}
\usepackage{listings}
\lstdefinelanguage{LF}{
  keywords={msec, sec, timer, startup, shutdown, state, main, actor, handler, reaction, preamble, target, reactor, composite, trigger, input, output, constructor, new, action, clock, logical, physical, after, import, from, private, public, method, time, interleaved, extends},
  keywordstyle=\color{black}\bfseries,
  ndkeywords={class, export, boolean, throw, implements, this, int, if, else, float},
  ndkeywordstyle=\color{darkgray}\bfseries,
  identifierstyle=\color{black},
  sensitive=false,
  comment=[l]{//},
  morecomment=[s]{/*}{*/},
  commentstyle=\color{black}\ttfamily,
  stringstyle=\color{stringColor}\ttfamily,
  morestring=[b]',
  morestring=[b]"
}
\begin{document}
        
\title{Strongly-Consistent \\
Distributed Discrete-event Systems}


\author{Peter Donovan, 
	Erling Jellum, 
	Byeonggil Jun, 
	Hokeun Kim, 
	Edward A. Lee,\\ 
	Shaokai Lin,  
	Marten Lohstroh, 
	Anirudh Rengarajan 
}


\markboth{Donvan et al.}%
{Strongly-Consistent Distributed Discrete-event Systems}


\maketitle
\setcounter{page}{1}
\thispagestyle{plain}

\begin{abstract}
Discrete-event (DE) systems are concurrent programs where components communicate via tagged events, where tags are drawn from a totally ordered set.  Reactors are an emerging model of computation based on DE and realized in the open-source coordination language Lingua Franca. Distributed DE (DDE) systems are DE systems where the components (reactors) communicate over networks. The prior art has required that for DDE systems with cycles, each cycle must contain at least one logical delay, where the tag of events is incremented.  Such delays, however, are not required by the elegant fixed-point semantics of DE. The only requirement is that the program be constructive, meaning it is free of causality cycles. This paper gives a way to coordinate the execution of DDE systems that can execute any constructive program, even one with zero-delay cycles. It provides a formal model that exposes exactly the information that must be shared across networks for such execution to be possible. Furthermore, it describes a concrete implementation that is an extension of the coordination mechanisms in Lingua Franca.
\end{abstract}


\section{Introduction}
Consistency is a measure of agreement about facts across different parts of a
system. But in order to arrive at any meaningful interpretation of the term
``agreement,'' we need to make assumptions about time. In discrete event
systems, it is typically assumed that all parts of the system have access to a
shared logical clock, and that events are tagged with a reading from that clock.
If the system is distributed, communication channels are used to distribute the
logical clock and relay tagged events across different parts of the system.
This is an effective way of accomplishing \emph{strong consistency} -- a situation where
each part of the system is able to make decisions based on information that all
parts of the system agree on.

Discrete-Event (DE) systems, and their distributed variant (DDE), have been
used for decades for modeling and simulation, including circuit and network
simulation, where time itself is modeled just like the rest of the system. More
recently, the (D)DE model has been adopted as a foundation upon which to build deterministic
concurrent software, where it takes the form of
Reactors~\cite{LohstrohEtAl:19:Reactors}, Logical Execution Time
(LET)~\cite{ErnstEtAl:18:LET}, or the Sparse Synchronous
model~\cite{EdwardsHui:20:SSM}. For these kinds of applications, which interact
directly with the physical world, time cannot always be treated as a purely
abstract, logical concept. In these systems, both a logical \emph{and} a
physical timeline are at play.

Considering the passage of logical time in conjunction with the passage of
physical time exposes that the communication required to achieve consistency has
a cost: it takes physical time. If decisions are held up by physical
communication delays in the system, this makes the system less available,
meaning that physical interactions are handled less quickly. The tolerance for
inconsistency and unavailability in a system is very application-dependent and
can vary across different parts of a system. The CAL
theorem~\cite{lee2021quantifying} shows us that consistency/availability
trade-offs can be navigated by inserting logical delays across communication
channels, effectively manipulating the tags of events. If a subsystem is
required to be strongly consistent, no logical delays can be inserted because it
has to allow for logically instantaneous communication. Yet, if a strongly
consistent subsystem has its distributed components organized such that there
exists a zero-delay cycle (ZDC) between those components, this poses a
significant coordination challenge. For this reason, most DDE frameworks reject
programs that have such ZDCs.


In this paper, we present a generalized DDE framework for the real-time
coordination of distributed DEs capable of handling ZDCs between nodes.
We formalize the coordination algorithms
required to ensure that data-determinism is preserved for all ``valid'' programs.
We extend Lingua Franca (LF), a recent coordination
language~\cite{LohstrohEtAl:21:Towards} with a DE semantics and ``preliminary''
support for distributed computing~\cite{BateniEtAl:23:Risk}. We show how the
coordination algorithm can preserve deterministic timed semantics for
distributed execution of all valid LF programs. A ``valid'' LF program is one
that has a constructive procedure that terminates at each tag in a finite number
of steps. LF has a constructive semantics~\cite{Berry:99:Constructive} and hence
rejects programs with causality cycles.

Our coordination strategy can handle ZDCs at the node level as long as there are no causality cycles. This enables elegant programming of \emph{strongly consistent}
distributed real-time systems. In practice today, ZDCs are avoided by
refactoring the model, requiring the ad-hoc introduction of logical
delays. This changes the model's semantics and might introduce errors
if the program depends on the logical simultaneity of certain events.
Our extension makes these contortions unnecessary.



The remainder of this paper is organized as follows.
Section~\ref{sec:related} discusses related work.
Section~\ref{sec:examples} gives simple examples that illustrate why it is attractive to support ZDCs.
Section~\ref{sec:tags} explains the time and tag model of Lingua Franca.
Section~\ref{sec:coordination} derives the information needed to ensure that events are handled in tag order.
Section~\ref{sec:TAG} describes the policy for allowing nodes to safely advance to a tag and process events with that tag.
Section~\ref{sec:blocking} gives implementation details for a simple realization of the policy.
Section~\ref{sec:evaluation} evaluates the implementation, showing that it has negligible impact on performance.
Finally, Section~\ref{sec:conclusions} draws conclusions.

\section{Related Work}\label{sec:related}
A DE model of computation (MoC), for our purposes here, is defined to be a concurrent system
where components react to tagged events in a tag order, where the tags
are drawn from a totally-ordered set~\cite{LeeSan:98:TaggedSignal}, usually representing time.
Similar timed models form the foundation of Simulink~\cite{karris2006introduction}, which is widely used in industry for building embedded software.
The DE MoC has a rigorous formal semantics that ensures data-determinism under well-defined assumptions~\cite{Baccelli:92:MaxPlus,Lee:99:DESemantics,Naundorf:00:Causal,Cataldo:06:Tetric,MatsikoudisLee:13:GenUltrametric}.
The DE MoC holds particular promise for real-time embedded software because of its integration of a model of time.
Lee and Zheng~\cite{LeeZheng:07:SRDECT} showed that the DE MoC can be
interpreted as a generalization of synchronous/reactive (SR) semantics (see also
Edwards and Hui~\cite{EdwardsHui:20:SSM}). The ``ticks'' of an SR semantics
become quantified by tags or timestamps, and each component reacts to tagged
inputs logically simultaneously, producing outputs with the same tag as the
inputs. Under this interpretation, a DE model inherits the elegant fixed-point
semantics of an SR model.

DDE systems for \emph{simulation} have a rich
history~\cite{Bryant:77:DE,Chandy:86:DDE,Jefferson:85:TimeWarp,Fujimoto:00:DistributedSimulation,ChenSzymanski:02:DE}.
A nice survey is given by Jafer, \textit{et al.}~\cite{JaferEtAl:13:DE}. Although most of
this work is quite old, research continues, motivated by network simulation and
digital twins, for example~\cite{GaoEtAl:23:DDE,HofmannEtAl:22:DDE}. However,
the DDE systems are rarely used for \emph{implementing} distributed real-time
systems. This is largely due to the challenges in preserving data-determinism
when nodes are communicating over possibly unreliable networks.

\section{Motivating Examples}\label{sec:examples}

\begin{figure*}
	\centering
	\includegraphics[width=1.35\columnwidth]{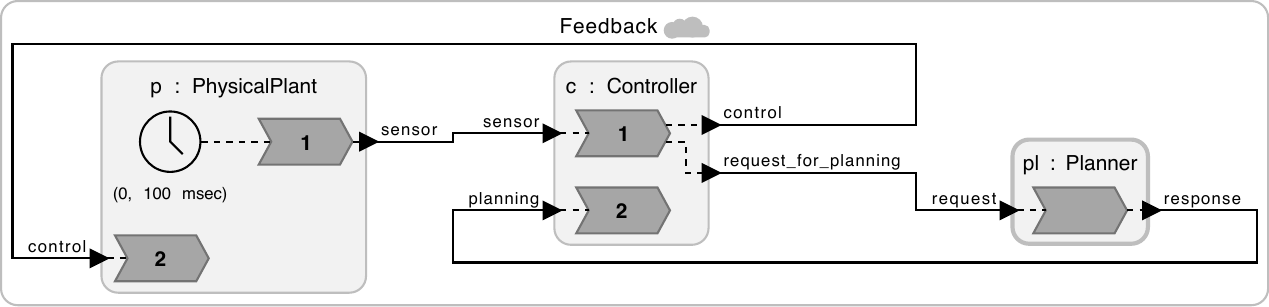}
	\caption{Control systems often have multiple coupled feedback loops that are usefully modeled as having zero delay.}
	\label{fig:Feedback}
\end{figure*}

\figurename~\ref{fig:Feedback} shows the (automatically generated) diagram for a Lingua Franca (LF) program realizing a feedback control system interacting with a physical plant.
The program consists of three reactor instances, named \code{p}, \code{c}, and \code{pl}, that send each other tagged messages via named ports.
Reactor \code{p} is an instance of a reactor class \code{PhysicalPlant}, which contains a timer and two reactions.
Reactions are represented in diagrams by the dark gray chevrons;
they consist of arbitrary code in a target language, which currently can be either C, C++, Python, Rust, or TypeScript.
Although the LF source code is textual, for the purposes of this paper, it is unnecessary to show this code.
The automatically generated diagrams are sufficient to convey all the required information.

In reactor \code{p}, the timer has an offset of 0 (meaning that it starts when the program starts) and a period of 100ms.
Reaction 1 is triggered by that timer and consists of arbitrary code in the target language.
Reaction 2 is triggered by an input on the \code{control} port, which will be a tagged message.

Reaction 1 in \code{p} could, for example, consist of C code that polls a sensor to produce sensor data on the output port named \code{sensor}.
Reaction 2 could, for example, consist of C code that drives an actuator.
Alternatively, these two reactions could interact with an external simulator or perform simulation themselves.
Here, the step size of such a simulation is fixed at 100ms, but it is easy in LF to vary the step size.

In \code{p}, if the timer event and the \code{control} input have the same tag (they are logically simultaneous), then reaction 1 will always be invoked before reaction 2, a key feature of LF that helps preserve determinism.
The order is defined by the lexical ordering of the reaction definitions in the source code and is reflected by their numbering in the diagram.
These reactions belong to the same reactor, and therefore have access to common state variables and/or the same physical plant.

Reactor \code{c} is an instance of \code{Controller}, which contains two reactions.
Reaction 1 is triggered by an input on its \code{sensor} port and (potentially) produces outputs on its \code{control} and \code{request\_for\_planning} ports.
If it produces such outputs, those outputs are \emph{logically simultaneous} with the inputs.
They have the same tag.
A control output, therefore, will feed back logically instantaneously to the physical plant.

\begin{figure*}
	\centering
	\includegraphics[width=1.45\columnwidth]{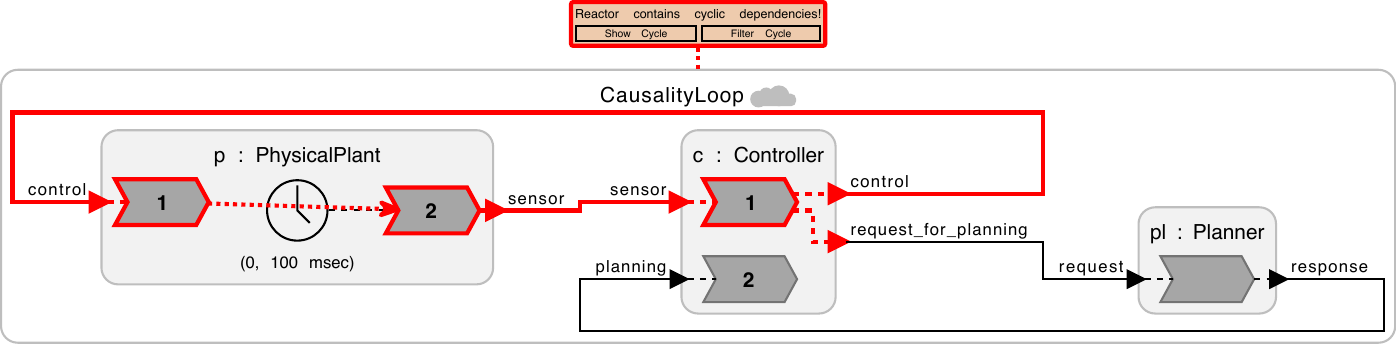}
	\caption{A causality loop is a cyclic dependency between logically simultaneous reactions.}
	\label{fig:CausalityLoop}
\end{figure*}

Notice that the ordering of reactions in the physical plant is important.
If reactions 1 and 2 were reversed, then we would have a causality loop, which is flagged by the LF tools as shown in \figurename~\ref{fig:CausalityLoop}.
The causality loop arises because, with this reversal, reaction 1 depends on the output produced by reaction 2,
but reaction 1 is required to be executed before reaction 2 at any tag where both are triggered.

The logically instantaneous feedback of \figurename~\ref{fig:Feedback} is a common pattern in control systems
and is repeated in the diagram in the interaction between the \code{Controller} and the \code{Planner}.
If the physical plant is simulated, e.g., using Simulink, the design would normally include an integrator in the physical plant simulation.
An integrator has the property that the output does not instantaneously depend on the input, and hence, it breaks the potential causality loop.

The pattern of \figurename~\ref{fig:Feedback} is not realizable in most DDE frameworks because there are ZDCs between the components. However, experiences in related systems suggest that it is possible to improve usability by accepting programs that intuitively should have a well-defined behavior, even if a conservative analysis of their component-level topology would indicate that they have a cyclic causal dependency~\cite{Berry:99:Constructive}.
Our generalization in this paper applies this insight to a DDE-like model of computation by enabling the execution of programs with ZDCs.
\peter{More citations here would be good. ``systems'' (plural) implies that this is not just one isolated example; if we cannot justify the use of the plural by providing more examples then that might not look good.}
\peter{This is a revised version of the commented-out paragraph which is immediately below in the TeX. I realize that it may be a bit of a stretch to go from synchronous languages to DDE frameworks, so it would be good to check that other people approve of the analogy.}

\begin{figure*}
	\centering
	\includegraphics[width=2\columnwidth]{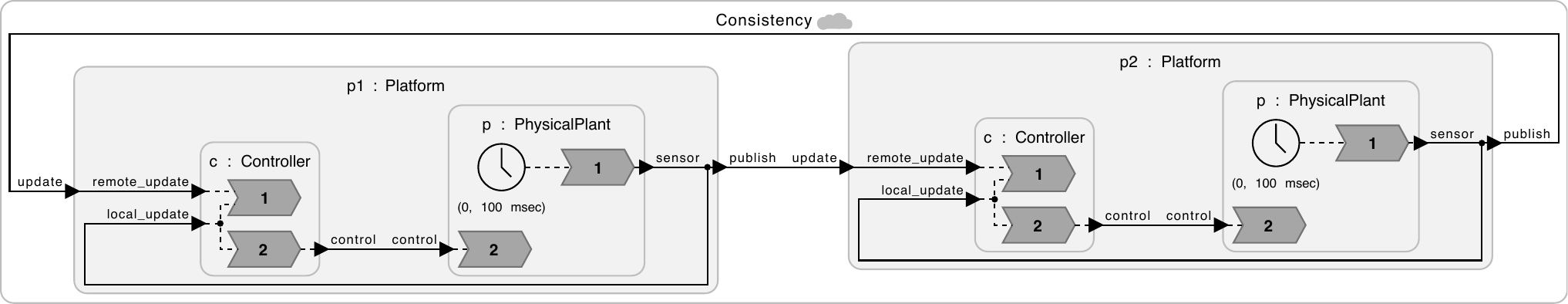}
	\caption{Distributed controllers whose control signal is based on a consensus.}
	\label{fig:Consistency}
\end{figure*}

A more interesting example is given in \figurename~\ref{fig:Consistency}, where two \code{Platform} instances each contain a physical plant interaction and a controller.
But now, we wish for the controller to make decisions based on a consensus across the platforms.
Such applications are often called multi-agent systems, and the pattern is common in robotics and vehicle control.
In this pattern, each \code{Controller} receives both a local update from its own physical plant interface
and an update from one or more remote platforms.
In reaction 1, the \code{Controller} calculates a consensus with confidence that the other platform will calculate
a similar consensus based on consistent information.

Notice the highly symmetric nature of \figurename~\ref{fig:Consistency}, where either platform can update common
information at any time, and both platforms can calculate a control signal with confidence that the other platform
calculates its control signal based on the same information. The CAL Theorem of Lee, \textit{et al.}~\cite{LeeEtAl:23:CAL_CPS}
shows that this \emph{strong consistency} comes at a price in availability that depends on the network latency (CAL
stands for consistency, availability, and latency). In this example, availability refers to the time it takes for the
\code{Controller} to reach the consensus state at which it can confidently issue a control signal. It is obvious in this
case that time will depend on how long it takes for these platforms to communicate. Moreover, Lee, \textit{et al.} show that
putting logical delays on the connections between these platforms introduces a measured amount of \emph{inconsistency},
which can reduce this amount of time, thereby improving availability. However, introducing logical delays must be done
very carefully to not change the semantics of the program. In the case of \figurename~\ref{fig:Consistency}, if a
logical delay is added to the connection from \code{p2} to \code{p1}, then at each multiple of 100ms, the \code{control} output of each \code{Controller} will be based on different information.  Without such a delay, even with simultaneous local and remote updates, both controllers have the same information and can, therefore, take consistent actions. With the delay, to ensure consistent actions, the same delay would have to be put on the connection from \code{p1} to \code{p2} and also on the internal connections from \code{sensor} to \code{local\_update}. This process is error-prone
and unnecessary if ZDCs are supported.

By extending the DDE execution framework to allow ZDCs, we enable the full gamut of design choices ranging from strong consistency (which requires ZDCs) to maximum availability (which can be achieved using LET, at the price of inconsistency). Note that consistency is defined with respect to logical time. This guarantee of strong consistency holds even in a granular notion of time, as we describe in the following section. Moreover, such a notion of time is fundamentally the right one for Lingua Franca because the deterministic semantics of Lingua Franca programs makes no reference to any notion of time that is even more granular.
\eal{I don't understand this comment.  Where is this coming from?  Why are we talking about granular time? OK to leave this in the submission, but we should revisit.}
\peter{Would appreciate at least one other person to check my changes to this paragraph.}
	
\section{Time and Tags}\label{sec:tags}

We make a sharp distinction between the logical and the physical timelines of a
distributed computation, as proposed by Lohstroh et.
al.~\cite{LohstrohEtAl:23:LogicalTime}.
We adopt the superdense
model of time used by LF~\cite{LohstrohEtAl:21:Towards} for logical time. Other models of time
accomplish the same
goals~\cite{BenvenisteEtAl:12:Nonstandard,CremonaEtAl:17:Hybrid}, but we do not
believe it would materially change our proposed DDE framework.

In LF, logical time is modeled with tags. A tag $g$ is a pair $(t, m) \in
	\mathbb{T} \times \mathbb{M}$, where $t \in \mathbb{T}$ is a 64-bit signed
integer obtained from the system clock, typically representing the number of
nanoseconds that have elapsed since 00:00:00 UTC on 1 January 1970. The least
and greatest numbers are reserved to represent {NEVER} and {FOREVER},
respectively, which are treated as infinities when performing arithmetic on time
values. The \emph{microstep} $m \in \mathbb{M}$ is an unsigned 32-bit integer
that allows a variable (such as an event at a port) to proceed through a
sequence of values without time elapsing. Given two tags $g_1 = (t_1, m_1)$ and
$g_2 = (t_2, m_2)$, $g_1 < g_2$ if and only if either $t_1 < t_2$ or $t_1 = t_2$
and $m_1 < m_2$. So, the set of tags forms a totally ordered set.

A port (input or output) may or may not have an event at any given tag $g$.
If it does not have an event, it is said to be \emph{absent} at that tag.
It can have at most one event at any $g$.
If two ports have events at the same tag, then these events are \emph{logically simultaneous}.
Output events produced by a reaction that is triggered by an input event are logically simultaneous with that input event (there is no microstep increment, as one finds in some DE languages, such as VHDL).

\subsection{Delays}\label{sec:delays}

The \code{after} keyword in Lingua Franca specifies a time value associated with a connection.
A message that travels over the connection has a tag $g_s = (t_s, m_s)$ at the sending end and a tag $g_r = (t_r, m_r)$ at the receiving end.
If there is no \code{after} on the connection, these two events are logically simultaneous.
An after delay is specified with $d \in \mathbb{T}$. If $d \ge 0$, then it specifies a tag increment across the connection;
the events are no longer logically simultaneous.
If $d = 0$, then the connection introduces a single microstep increment.
Since a specified after delay always satisfies $d \ge 0$, we can model
the lack of an after delay with any $d < 0$.
Formally, then, the tag at the receiving end is given by a function $D$ defined by,
\begin{equation}\label{eq:D}
	\begin{aligned}
		D(g_s, d) =
		D((t_s, m_s), d) =
		(t_r, m_r) = \\
		\begin{cases}
			(t_s, m_s),         & \text{if } d < 0                      \\
			(t_s, m_s + 1),     & \text{if } d = 0                      \\
			(t_s + d, 0)        & \text{if } d > 0                      \\
			\text{FOREVER\_TAG} & \text{if } t_s \ge \text{FOREVER} - d \\
			\text{FOREVER\_TAG} & \text{if } d = \text{FOREVER}
		\end{cases}
	\end{aligned}
\end{equation}
Notice that when an after delay with $d > 0$ is given, the microstep at the receiving end is reset to zero.
We will not defend that choice here, but rather just accept it as the behavior exhibited by Lingua Franca.
In the last two conditions, FOREVER\_TAG is the largest representable tag.
The last condition means we can model the lack of connection with $d =$ FOREVER.
The second to last condition models the situation where the time value overflows.

Consider a network of reactors.
Let $D_{ij}$ be the \emph{least} \code{after} delay over all connections from reactor $j$ to $i$.
If there are no connections from $j$ to $i$, let $D_{ij} =$ {FOREVER}, the largest possible time.
If $i=j$, let $D_{ij} =$ {NEVER}, the least (and negative) time.

If at a tag $g_j$, $j$ sends a message to $i$, then the tag $g_i$ at the receiving end is at least as large as $D(g_j, D_{ij})$.
Equivalently,
\[
	g_i \ge g_j + D((0,0), D_{ij}),
\]
where the addition of tags is element-wise.
We call $D((0,0), D_{ij})$ the \textbf{minimum tag increment} from $j$ to $i$.

So far, we are only talking about direct connections between reactors.
A message from one reactor, however, can trigger many other messages downstream, even an infinite number if there are cycles.
We will be interested in the minimum possible tag increments along paths that may pass through several intermediate reactors.

\begin{figure}
	\centering
	\subfloat[An example where $D_{CA}$ is $(10 ms, 0)$.]{\includegraphics[width=0.6\columnwidth]{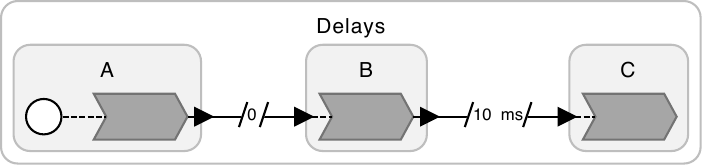}\label{fig:Delay1}}\\
	\subfloat[An example where $D_{CA}$ is $(10 ms, 1)$.]{\includegraphics[width=0.6\columnwidth]{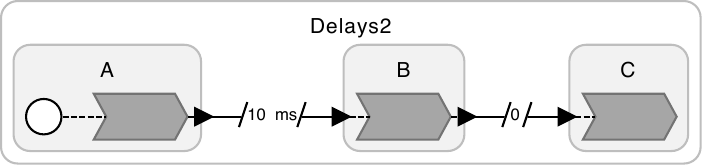}\label{fig:Delay2}}\\
	\subfloat[An example with a path including a logical action.]{\includegraphics[width=0.8\columnwidth]{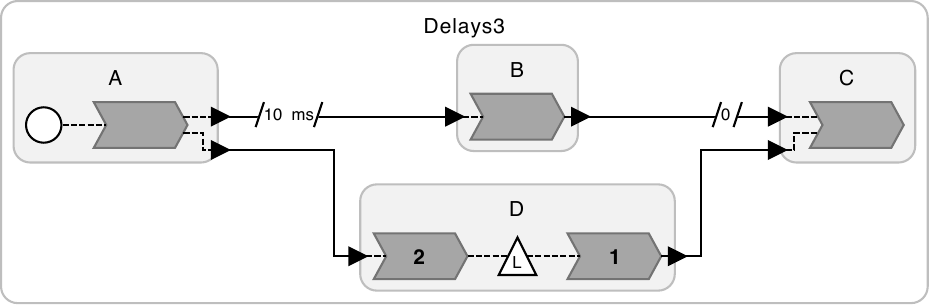}\label{fig:Delay3}}
	\caption{Example LF programs with \code{after} delays.}
\end{figure}

\IEEEpubidadjcol{} 
Let $\mathcal{D}_{ij}$ denote the \textbf{minimum tag increment} along any \emph{path} from node $j$ to node $i$.
This tag increment is calculated using the $D$ function as above, applied in the order in which connections appear along the path.
For example, consider a program in \figurename~\ref{fig:Delay1}.
The connection from $A$ to $B$ has an \code{after} delay of 0 (implying a one microstep delay),
and the connection from $B$ to $C$ has an \code{after} delay of 10ms.
Hence, the minimum tag increment from $A$ to $C$ is
\[
	\mathcal{D}_{CA} = D(D((0,0), 0), 10000000) = (10000000, 0).
\]
We start with $A$ at a tag $(0,0)$ and evaluate the $D$ function, first along the
connection from $A$ to $B$, and then along the connection
from $B$ to $C$.

Note that the $D$ function is neither commutative nor associative, so when finding the total delay of a path consisting of several connections, it is necessary to apply the function in the order of the connections.
For example, the example in \figurename~\ref{fig:Delay2} yields a minimum tag increment of
\[
	\mathcal{D}_{CA} = D(D((0,0), 10000000), 0) = (10000000, 1).
\]
If both delays were 0, then we would get  $\mathcal{D}_{CA} = (0,2)$.

Note that in both cases \figurename~\ref{fig:Delay1} and \figurename~\ref{fig:Delay2}, there is no path from $C$ to $A$, so $\mathcal{D}_{AC} =$ FOREVER\_TAG.
Moreover, there is no path from $A$ to itself, so we define $\mathcal{D}_{AA} =$ FOREVER\_TAG.

The minimum tag increment is conservative.
It does not take into account causality interfaces~\cite{ZhouLee:08:CausalityInterfaces}.
Causality interfaces are algebraic models that express how each output port depends on each input port.
If an input at a tag $g$ can cause an output at the tag $g$, then there is a \textbf{direct feedthrough} path
from that input port to that output port. If an input at a tag $g$ cannot cause an output before the tag $g + d$, then
$d$ represents a \textbf{minimum delay} from that input port to that output port.

Hence, for example, the program in \figurename~\ref{fig:Delay3}, $\mathcal{D}_{CA} = (0,0)$ even
though it is not possible for an output from $A$ to trigger an input to $C$ at the
same tag.
The white triangle in $D$ represents a \emph{logical
	action}, used in Lingua Franca for one reaction to schedule the invocation of
another reaction at a later tag. In this example, reaction 2 schedules a later
invocation of reaction 1, so the resulting output will never be logically
simultaneous with the input. If we were using causality interfaces, we would
capture this fact. Our conservative definition ignores this fact and assumes
that there might be direct feedthrough from the input to D to its output.

It turns out that this conservative definition is almost sufficient to get correct coordination, as we will see below.
It is sufficient to determine the tag to which a node may commit (explained in Section~\ref{sec:TAG}), but additional information is
used to prevent premature blocking on unknown input port statuses (explained in Section~\ref{sec:blocking}).
In this example, the output from D that is the result of an input has a minimum tag increment of $D((0,0), \text{min\_delay})$, where min\_delay is the minimum delay parameter of the logical action.
The default minimum delay is 0, resulting in a minimum tag increment of $(0,1)$.
If we were to use such information, then we would get
\[
	\mathcal{D}_{CA} = \min((10000000, 1), D((0,0), \text{min\_delay})).
\]
This is greater than or equal to our more conservative result $\mathcal{D}_{CA} = (0,0)$.
That result is ``more conservative'' because we use  $\mathcal{D}_{CA}$ to determine the earliest possible tags of future messages received by C that are caused by events at A, and the error here results in an earlier estimate rather than a later estimate.
A later estimate could result in C seeing events out of the tag order, which would violate the semantics of LF.

\section{Centralized Coordination}\label{sec:coordination}

Prior to this work, Lingua Franca has provided two preliminary DDE execution frameworks that handle coordination
to ensure that components see events in the tag order~\cite{BateniEtAl:23:Risk}.
The more conservative coordinator, called the ``centralized coordinator,'' is based loosely on the High-Level Architecture (HLA)~\cite{KuhlEtAl:99:HLA}.
A second coordinator, called the ``decentralized coordinator,'' is based on PTIDES~\cite{Zhao:07:PTIDES} (using techniques similar to those in Google Spanner~\cite{CorbettEtAl:12:Spanner}).
In this paper, we extend the centralized coordinator to support ZDCs (as long as they are constructive).
Although our centralized coordinator is implemented as a single centralized process, there is no fundamental reason that its functionality could not be distributed among the other components, yielding a more decentralized realization of the same semantics.  This paper, in fact, clarifies exactly what information needs to be shared across a distributed system in order to make such coordination possible.

\begin{table}
	\centering
	\begin{tabular}{lll}
		Name       & Description                   & Direction          \\
		\toprule
		ABS$_{ij}$ & Absent                        & $j$ to $i$ via RTI \\
		LTC$_i$    & Latest Tag Complete          & $i$ to RTI         \\
		MSG$_{ij}$ & Message                       & $j$ to $i$ via RTI \\
		NES$_i$    & Neighbor structure            & $i$ to RTI         \\
		NET$_i$    & Next Event Tag                & $i$ to RTI         \\
		PTAG$_i$   & Provisional Tag Advance Grant & RTI to $i$         \\
		TAG$_i$    & Tag Advance Grant             & RTI to $i$         \\
		\bottomrule
	\end{tabular}
	\vspace*{3pt} 
	\caption{Types of messages and signals exchanged between nodes and the RTI.\label{tab:message_types}}
\end{table}

\subsection{Message and Signal Types}\label{sec:msgtypes}

Message and signal types needed for coordinating the distributed execution are
summarized in Table \ref{tab:message_types}. There are two types of entities in
the coordination, nodes and the Run-Time Infrastructure (RTI). A node (also called a federate) is
a top-level reactor that is part of a distributed execution (also known as a
federation). The RTI is a central coordinator that has a direct communication
link with every node. We use the term ``message'' for application-level
messages that go from one node to another. We use the term ``signal'' for
control messages exchanged with the RTI.

Most of these messages and signals carry a tag as part of the payload. Let
$G(M_i)$ denote the tag carried by the message or signal $M_i$. For example,
$G(\text{LTC}_i)$ is the tag on an LTC$_i$ signal, which tells the RTI that node
$i$ has completed processing tag $G(\text{LTC}_i)$. This means that the RTI has seen
all messages from $i$ destined to arrive at any other node with tags less than or equal to
$G(\text{LTC}_i)$ plus the after delay of the connection between the nodes.

When the RTI receives a message MSG$_{ij}$ from node $j$, it will forward that
message to node $i$. That message has two tags associated with it, a \textbf{source tag} $g_j$,
the tag at source node $j$, and $g_i$, the \textbf{destination tag} at node $i$.
The destination tag $g_i$ is the source tag $g_j$ adjusted by any \code{after}
delay on the connection from $j$ to $i$. I.e., $g_i = D(g_j, d)$, where $d$ is
the \code{after} delay on the connection from $j$ to $i$, and $D$ is the
function defined in equation (\ref{eq:D}). Only the destination tag is carried
as a payload along with the message, so $g_i = G(\text{MSG}_{ij})$. A message
MSG$_{ij}$ can occur only if there is a connection from an output port of
node $j$ to an input port of node $i$. 

An ABS$_{ij}$ is similar to MSG$_{ij}$ except that it conveys to node $i$ that it will never in the future
receive a message from $j$ with a tag less than or equal to $G(\text{ABS}_{ij})$.
ABS signals are used only when there are ZDCs, as we will see below.

A NET$_i$ signal from node $i$ is a promise that, as of the time at which it
sends this signal to the RTI, it will not produce any further messages to other
nodes with tags less than $G(\text{NET}_i)$, except in response to receiving a
message from another node $j$ with a tag $G(\text{MSG}_{ij}) < G(\text{NET}_i)$.

A TAG$_i$ signal is sent from the RTI to node $i$ when it is safe for node $i$ to assume it has received
all MSG$_{ij}$ and ABS$_{ij}$ with tags less than or equal to $G(\text{TAG}_i)$.
This informs node $i$ that it can proceed to process all such messages.
A PTAG$_i$ signal is sent from the RTI to node $i$ when it is safe for node $i$ to assume it has received
all MSG$_{ij}$ and ABS$_{ij}$ with tags \emph{strictly less than} $G(\text{PTAG}_i)$.
This informs node $i$ that it can safely advance its local tag to $G(\text{PTAG}_i)$ (it will not later see messages with earlier tags), but it is not yet safe to assume it has seen messages with tags \emph{equal} to $G(\text{PTAG}_i)$.
PTAG signals, like ABS signals, are used only when there are ZDCs. In
Section~\ref{sec:blocking} introduces a strategy that partially
executes logical tags at a node upon receiving a PTAG signal.

In the following section, we introduce a coordination algorithm specifying how the RTI
should send TAG and PTAG signals in response to LTC and NET signals and MSG and
ABS messages.

\subsection{Topology Information}\label{sec:topology}

At the start of execution, each node informs the RTI of the connection topology.
Each node is identified by a unique ID and must know the IDs of its upstream
and downstream nodes.

Recall that $D_{ij}$ is the minimum delay on connections from an output port of node $j$ to an input port of $i$.
If there is no connection from $j$ to $i$, then $D_{ij}$ is {FOREVER}.
If there are connections, but no connection from $j$ to $i$ has an \code{after} delay, then $D_{ij}$ is {NEVER}.
Similarly, each node can be considered to be connected to itself with no after delay, so $D_{ii}$ is {NEVER}.

Each node $i$ sends to the RTI a NES$_i$ signal that lists the IDs of the nodes that are upstream and downstream of it, and for each upstream node $j$, the minimum delay $D_{ij}$ from $j$ to $i$.
This will be the sum total of the information that the RTI has about the connection topology.
For distributed programs that support dynamic topologies, this information will need to be updated at runtime.

\subsection{Assumptions}\label{sec:assumptions}


First, a node $j$ can only produce an output in response to:
\begin{enumerate}[label=\Alph*.]
	\item\label{it:causeMSG} An input MSG or ABS from the network.
	\item\label{it:causeEvent} An event on its event queue.
	\item\label{it:causePhysical} A physical action.
\end{enumerate}
We call these the \textbf{causes} for messages $\text{MSG}_{ij}$ from $j$ to $i$.

A \emph{physical action} is a Lingua Franca concept~\cite{LohstrohEtAl:21:Towards}.
It is a mechanism that enables injection into an executing program of tagged events that are triggered by some external event, such as sensor data delivered via an interrupt.
The timestamp in the tag of the resulting event is determined by the local physical clock at the node where the event is injected.
When such physical actions are present, a node cannot process tagged events with timestamps greater than its local physical clock because, if it did, then a physical action may inject an event that is in the past relative to events that have been processed.
As a consequence, logical time (the timestamps of events being processed) is always behind physical time;
we can say, ``logical time chases physical time.''
As a consequence, because of the definition of $\text{NET}$, if node $i$ has a physical action, it never sends $\text{NET}_i$ with $g(\text{NET}_i) \ge T$, where $T$ its physical time.

In addition, we need to make the following \textbf{assumption} about the execution:
	      All messages and signals from each node $i$ to RTI and from the RTI to $i$ are delivered reliably and in order.
	      (This can be realized using TCP.)

\subsection{Earliest Future Incoming Message Tag}\label{sec:EIMT}

When the RTI receives a $\text{NET}_i$ from $i$, it must decide whether to grant a
$\text{TAG}_i$, a $\text{PTAG}_i$, or neither. To do this, it needs a lower bound $B_i$ on the
tags of messages that $i$ may receive in the future. We call $B_i$ the
\textbf{earliest (future) incoming message tag} (\textbf{EIMT}) for node $i$.

If the EIMT $B_i > G(\text{NET}_i)$, then the RTI can grant a TAG$_i$ with $G(\text{TAG}_i) =
	G(\text{NET}_i)$. If $B_i = G(\text{NET}_i)$, then the RTI can grant a $\text{PTAG}_i$ with
$G(\text{PTAG}_i) = G(\text{NET}_i)$, although it may choose not to grant the $\text{PTAG}_i$  if $i$
is not in part of a ZDC (see Section~\ref{sec:TAG} below). If  $B_i < G(\text{NET}_i)$,
then the RTI must grant neither. To determine $B_i$, the RTI needs to use the
information it has about the \emph{other} nodes that may send messages to $i$.
Specifically, for each node $j$, it needs to find a lower bound on the tags of
any message that $j$ may send to $i$. In this section, we explain how to find
this lower bound.

Consider a message $\text{MSG}_{ij}$ sent from node $j$ to node $i$.
Recall that $D_{ij}$ is the minimum \code{after} delay over all connections from $j$ to $i$.
Then the tag $g_i$ of the message at the receiving node will satisfy
\[
	g_i = G(\text{MSG}_{ij}) \ge D(g_j, D_{ij}),
\]
where $g_j$ is the tag at the sending node.

Note that the tags $G(\text{MSG}_{ij})$ on a sequence of messages from $j$ to $i$ may
not be monotonically increasing. It is possible for node $j$ to have more than
one connection to node $i$ (on different ports), and these connections may have
arbitrary \texttt{after} delays. Hence, the RTI may see a MSG$_{ij}$ with destination tag
$g_i$ and later see a similar message with an earlier destination tag.

For each node $i$, the RTI maintains a variable $N_i$ that is the tag of the
most recently received $\text{NET}_i$ signal from $i$. If no $\text{NET}_i$ has been received,
then $N_i$ is the start tag of the execution, a pessimistic guess for the
earliest tag at which $i$ might produce a message.

Once the RTI sees a message $\text{MSG}_{ij}$, then it knows that $i$ will need to
execute at the tag $G(\text{MSG}_{ij})$ even if it has not yet received a NET$_i$ with
$G(\text{NET}_i) = G(\text{MSG}_{ij})$. Thus, the RTI updates $N_i$ to $G(\text{MSG}_{ij})$ if
$G(\text{MSG}_{ij})$ is earlier than $N_i$. Only when the RTI receives an LTC$_i$ with
a tag greater than or equal to $G(\text{MSG}_{ij})$ can it be sure that node $i$ has
finished processing the message and produced any outgoing messages in response
to it.

For each node $i$, the RTI also maintains a sorted queue $Q_i$ of tags
$G(\text{MSG}_{ij})$ of messages destined to $i$. Intuitively, $Q_i$ is a sorted record
of all messages that are in flight to $i$ but have not yet been processed by
$i$, as far as the RTI knows. When it receives an LTC$_i$, it pops off
$Q_i$ all tags less than or equal to $G(\text{LTC}_i)$ and discards them because
it now knows that $i$ has processed all those messages. Consequently, the RTI
has also seen all messages $\text{MSG}_{ki}$ from $i$ to any other node $k$ with
\emph{source} tag equal or less than $G(\text{NET}_i)$. Let $H(Q_i)$ be the head
of $Q_i$ (the least tag on the queue) or {FOREVER} if $Q_i$ is empty.
The necessity of the queue is explained in Section~\ref{sec:in-transit}.


Define $E_i$ to be a lower bound on the \emph{source} tag of any future messages (messages not yet seen by the RTI) that node $i$ may produce. 
If the RTI can calculate $E_i$ for each node $i$, then it can also calculate the EIMT $B_i$, a lower bound on the tags of messages that node $i$ may receive in the future.  This will be
\begin{equation}\label{eq:Bp}
	B_i = \min_{j \in U_i} D(E_j, D_{ij}),
\end{equation}
where $U(i)$ is the set of nodes immediately upstream of $i$, and $D$ is the delay function given in Section \ref{sec:delays}.
As explained above, the RTI can use this to decide how to respond (or whether to respond) to a $\text{NET}_i$ signal.

Future messages produced by a node $i$ can only have one of two causes: they result from either an event originating within node $i$ or a message sent to $i$ from some other node.
Hence, we can express $E_j$ from Eq.~\ref{eq:Bp} in terms of $B_j$,
i.e., the EIMT of an upstream node $j$, and other information the RTI has:
\begin{equation} \label{eq:E}
	E_j = \min(N_j, H(Q_j), B_j).
\end{equation}
Hence, the EIMT of node $i$ can be given by
\begin{equation} \label{eq:B}
	B_i = \min_{j \in U_i} D( \min(N_j, H(Q_j), B_j), D_{ij}).
\end{equation}
The RTI will never in the future see a message $\text{MSG}_{ij}$ destined for $i$ with a tag $G(\text{MSG}_{ij}) < B_i$.

To understand why this is a lower bound, consider each term of (\ref{eq:E}), $N_j$, $H(Q_j)$, and $B_j$:

\noindent
(a)
Recall that $N_j$ is the tag of the most recently received $\text{NET}_j$ from $j$, meaning that $j$ may produce a message at $N_j$.
No matter the RTI has seen such a message or not, it must assume such a message is possible.
Notice, however, that there is no assurance that $\text{NET}_j$ signals have monotonically increasing tags.
In fact, the RTI may have seen a signal with a tag $G(\text{NET}_j) = FOREVER$ (or a large value equal to the timeout tag), indicating that $j$ does not have any internal events that could result in a message. 
But $j$ may subsequently (after having sent FOREVER) receive a message that causes it to produce a message to $i$ with an earlier tag.
In that case, however, the tag of that message will already be accounted for in $B_j$ (or $Q_j$ if the RTI has seen that message).
$B_j$ is the tag of the earliest possible incoming message to $j$, as far as the RTI knows, and it will not be updated until the RTI is sure that all messages with tags earlier than $B_j$ have been delivered to $j$ and $j$ has acknowledged receiving it with a $\text{LTC}_j$.

\noindent
(b)
$Q_j$ contains the tags of messages that might be in flight to $j$, each of which could cause $j$ to produce a message in response.

\noindent
(c)
By the definition of $B_j$, $j$ will not receive a message with a tag earlier than $B_j$, so if such a message triggers a message to $i$, that message cannot have a \emph{source} tag less than $B_j$.

This accounts for all the possible causes of future messages from $j$.
The validity of this conclusion depends on the fact that the channel from $j$ to the RTI has reliable in-order delivery of messages and signals.
Specifically, a message MSG$_{ij}$ from $j$ with a source tag $g_j$ will always be preceded by a NET$_j$ with $G(\text{NET}_j) = g_j$, and no other NET$_j$ signal will arrive in between.
The $H(Q_j)$ accounts for the possibility that the most recent NET$_j$ corresponds to a \emph{future} message, one that the RTI has not yet seen, but the RTI is aware of a message in flight towards $j$.
The final term accounts for the possibility that the RTI has not yet seen relevant messages destined towards $j$, but those messages have to originate at upstream nodes.

Note that equation (\ref{eq:B}) is circular. If there is a path upstream from $i$ back to itself, then $B_i$ depends on itself.
Fortunately, this does not create a problem, as we will see.

\subsection{Two-Node Case}

Consider a system with just two nodes, $i,j \in \{0, 1\}$.
Then, using the definitions in Section \ref{sec:delays}, equation (\ref{eq:B}) reduces to
\[
	\begin{split}
		B_0 & = D(\min(N_1, H(Q_1), B_1), D_{01}) \\
		B_1 & = D(\min(N_0, H(Q_0), B_0), D_{10})
	\end{split}
\]
This is the situation when 1 and 2 are not part of a ZDC.
Combining these, we see that the EIMT for node 0 is
\begin{equation}\label{eq:2B}
	\begin{aligned}
		B_0 = D( & \min(N_1, H(Q_1),         \\
		         & D(\min(N_0, H(Q_0), B_0), D_{10})), D_{01} ).
	\end{aligned}
\end{equation}
Notice that $B_0$ appears on both sides.
The $D$ function is monotonic, so we can exchange $\min$ and $D$ to get
\[
	\begin{split}
		B_0 = \min( & D(N_1,  D_{01}),  \\
		&  D(H(Q_1), D_{01}),  \\
		&  D(D(N_0, D_{10}), D_{01}), \\
		&  D(D(H(Q_0), D_{10}), D_{01}), \\
		&  D(D(B_0, D_{10}), D_{01}) ).
	\end{split}
\]
If either $D_{10} >= 0$ or  $D_{01} >= 0$, then
the last of these cases cannot be the minimum.
That is, it cannot be that
$
	B_0 = D(D(B_0, D_{10}), D_{01}).
$
Hence, if either $D_{10} >= 0$ or  $D_{01} >= 0$, then,
\begin{equation}\label{eq:delay}
	\begin{split}
		B_0 = \min( & D(N_1,  D_{01}),  \\
		&  D(H(Q_1), D_{01}),  \\
		&  D(D(N_0, D_{10}), D_{01}), \\
		&  D(D(H(Q_0), D_{10}), D_{01})).
	\end{split}
\end{equation}
This is the situation where either there is no cycle between nodes 0 and 1 (one
of $D_{01}$ or $D_{10}$ is FOREVER), or there is a cycle but with a logical
delay.

In the case that there is a ZDC, $D_{01}$ and $D_{10}$ are
both negative and
\begin{equation}\label{eq:nodelay}
	B_0 = \min(
	N_1,
	H(Q_1),
	N_0,
	H(Q_0),
	B_0
	).
\end{equation}
Let
$
	X = \min(
	N_1,
	H(Q_1),
	N_0,
	H(Q_0)
	).
$
Then it is clear that any $B_0 \le X$ satisfies (\ref{eq:nodelay}).
However, this conclusion is unnecessarily weak, and if we take causality into account, we can instead
make the stronger conclusion that $B_0 = X$.
This is because,
based on the causes listed in Section \ref{sec:assumptions}, the appearance of $B_0$ in
the minimization would have to correspond to a case when the cause of some event $e$ at a tag $B_0$ is $e$ itself. Lingua Franca forbids causality loops, and
hence such a case is impossible and should not be considered. Hence, (\ref{eq:delay}) is a general solution,
applicable for any delays $D_{01}$ and $D_{10}$.

\erj{I think we must explain why "B0 less than X" indicates that we have an event without
	one of those causes and the link to causality cycles}
\peter{Update: Slightly adjusted the wording. In light of the edits, it would seem that the presentation would be clearer if we had excluded $B_0$ from the beginning, and simply written, $B_i = \min_{j \in U_i} D( \min(N_j, H(Q_j), (B_j \,\text{if}\, j \neq i \,\text{else}\, \infty)), D_{ij}).$.}
\subsection{$N$-Node Case}\label{sec:algorithm}

Equation (\ref{eq:delay}) can be generalized to an arbitrary set $N$ of nodes using the path delay $\mathcal{D}_{ij}$ defined in Section \ref{sec:delays}.
For all $i \in N$, the EIMT is
\begin{equation}\label{eq:alg}
	B_i = \min_{j \in N} (N_j + \mathcal{D}_{ij}, H(Q_j) + \mathcal{D}_{ij}).
\end{equation}
Equivalently,
\begin{equation}\label{eq:alg2}
	B_i = \min_{j \in N} ( \min(N_j, H(Q_j)) + \mathcal{D}_{ij}).
\end{equation}
Equation (\ref{eq:alg2}) suggests an algorithm.
For each node $i$ and each node $j$ that is upstream of $i$ along at least one path with no repetitions of nodes, we can calculate $\mathcal{D}_{ij}$, which is the shortest path, using Dijkstra's algorithm.
This calculation needs to be done only once if the topology does not change during runtime, and recalculated only when the topology changes.
Then, when the RTI receives a $\text{NET}_i$ signal, it can grant a TAG$_i$ with $G(\text{TAG}_i) = G(\text{NET}_i)$ if $B_i > G(\text{NET}_i)$.
If $B_i = G(\text{NET}_i)$, then there may be a ZDC, and the RTI can only grant a $\text{PTAG}_i$ with $G(\text{PTAG}_i) = G(\text{NET}_i)$.
If $B_i < G(\text{NET}_i)$, then it cannot grant anything.
The decision of whether to grant a TAG$_i$ or $\text{PTAG}_i$ should be revisited whenever a $\text{NET}_j$ is received from any upstream node $j$.

\section{Tag Advance Grants}\label{sec:TAG}

When a node becomes idle (and at startup), having processed all the events at a
particular tag $g$, it enters a state where it wishes to advance its tag to some
$g' > g$. If it has no events on its event queue, then $g'$ is either
$(\text{TIMEOUT}, 0)$, where TIMEOUT is the logical stop time, or FOREVER\_TAG,
if no logical stop time is specified. The node sends a NET$_i$
signal to the RTI with $G(\text{NET}_i) = g'$. If the node has upstream nodes, it then
waits for a response from the RTI. The response will come in the form of a
TAG$_i$ or $\text{PTAG}_i$ with a tag greater than $g$ and no greater than $g'$. If
node $i$ has no upstream nodes, then it does not wait, and instead acts as if it
has already received a TAG$_i$ with $G(TAG_i) = g'$.

Recall that a TAG$_i$ signal from the RTI is a guarantee to node $i$ that it has seen all
network input messages with tags less than or equal to $G(\text{TAG}_i)$, and thus can
safely advance its logical tag to TAG$_i$.
Also, recall that a  $\text{PTAG}_i$ signal from the RTI is a guarantee to node $i$ that it has seen all
network input messages with tags less than $G(\text{PTAG}_i)$.

Notice that the difference between these is subtle but important. In the case
of a TAG$_i$, the node can assume that any input port that has not received a
message with a tag $G(\text{TAG}_i)$ is \textbf{absent} at the tag $G(\text{TAG}_i)$.

In the case of a PTAG$_i$ signal, however, it cannot make any such assumption.
The state of each input port is \textbf{unknown} at a tag $G(\text{PTAG}_i)$ unless it
has already received a message for that port with that tag (or a greater tag).
As the node executes reactions at a tag $G(\text{PTAG}_i)$, it must block any reaction
that depends on an unknown input port until that port becomes \textbf{known} (to have a message or to be absent).  There are several ways that an
unknown port may become known:
\begin{enumerate}
	\item The port receives a message $\text{MSG}_{ij}$ with a tag $G(\text{MSG}_{ij}) \ge G(\text{PTAG}_i)$. If greater, then the port becomes known to be absent.
	\item The port receives an $\text{ABS}_{ij}$ message with $G(\text{ABS}_{ij}) \ge G(\text{PTAG}_i)$. At this point, the port is known to be absent.
	\item The node receives a TAG$_i$ signal from the RTI with $G(\text{TAG}_i) \ge G(\text{PTAG}_i)$.
	\item The node can infer from the interconnection topology that the RTI would not have granted it the $\text{PTAG}_i$ without having forwarded all possible messages destined for the port with $G(\text{MSG}_{ij}) \le G(\text{PTAG}_i)$.
\end{enumerate}
The last of these is the most subtle and deserves some attention.
First, we need to examine the policy that the RTI uses to send TAG and PTAG signals.

\subsection{Policy for Granting Tag Advances}
Let $C_i$ be the tag of the most recent LTC$_i$ signal that the RTI has seen from node $i$, or NEVER\_TAG if there has been no such signal.
When node $i$ is ready to make progress, it will send a NET$_i$ signal to the RTI with $G(\text{NET}_i) > C_i$.
When the RTI receives such a NET$_i$ from node $i$, it can do one of three things:
\begin{enumerate}
	\item It can respond with a TAG$_i$.
	\item It can respond with a $\text{PTAG}_i$.
	\item It can do nothing and wait for more information. This will block progress at node $i$.
\end{enumerate}
To determine which of these to do, it should use the algorithm of Section~\ref{sec:algorithm} to calculate $B_i$,
the earliest (future) incoming message tag (EIMT).
If the system is behaving correctly, then it should always be true that If $B_i > C_i$; otherwise an error has occurred.
Assuming such an error does not occur, there are several possible outcomes:
\begin{enumerate}
	\item If $B_i > G(\text{NET}_i)$, then the RTI can issue a TAG$_i$ with $G(\text{TAG}_i) = G(\text{NET}_i)$.
	\item If $B_i = G(\text{NET}_i)$, then the RTI has two choices:
	      \begin{enumerate}[label=\alph*.]
		      \item The RTI does nothing, waiting for further information.
		      \item The RTI issues a $\text{PTAG}_i$ with $G(\text{PTAG}_i) = G(\text{NET}_i)$.
	      \end{enumerate}
	\item If $B_i < G(\text{NET}_i)$, the RTI \emph{must} do nothing and wait for further information.
\end{enumerate}

There are two relatively simple strategies that an RTI design could follow, but both of them have serious disadvantages.
First, it could always choose option 2a above.
However, this will result in a large number of ABS messages becoming necessary.
Every time a node $i$ completes execution at some tag $g$, before sending its $\text{LTC}_i$ signal, it will need to send $\text{ABS}_{ji}$ messages to each downstream node $j$ on every output port where it did not actually produce a $\text{MSG}_{ji}$.
This will clog the network with absent messages.
Second, it could always choose option 2b above. However, this will result in an
inability to support ZDCs.


\subsection{Coordination Strategy}
There are two questions to resolve.
First, faced with $B_i = G(\text{NET}_i)$ (condition 2 above), should the RTI choose 2a or 2b?
Second, when the RTI chooses either 2a or 3 (it waits), then when does the RTI finally respond to $\text{NET}_i$?

For the first question, our strategy is to minimize the use of $\text{PTAG}$ and $\text{ABS}$; using them only when not using them would result in a deadlock.
Hence, when $B_i = G(\text{NET}_i)$, it becomes important which node(s) upstream of $i$ are the ones that may send messages with tags $B_i$.
If any one of those nodes is in a ZDC with $i$, then the RTI will choose 2b and issue a $\text{PTAG}_i$.
Otherwise, it will choose 2a.
It is safe to choose 2a because if all upstream nodes $j$ that may produce a message $\text{MSG}_{ij}$ with $G(\text{MSG}_{ij}) = B_i = G(\text{NET}_i)$ are not in a ZDC with $i$, then they will be able to progress to the tag $B_i$ without waiting for $i$ to progress to the tag $B_i$, and, hence, not causing a deadlock.

The second question is when the RTI finally responds to a $\text{NET}_i$ when it has chosen to wait.
Once the RTI receives an $\text{LTC}_j$ with $G(\text{LTC}_j) \ge \text{NET}_i$ from each $j$ that is upstream of $i$, then the RTI can issue a TAG$_i$ with TAG$_i = \text{NET}_i$ (or a possibly even larger tag, the minimum of all the $g(\text{LTC}_j)$).
Hence, upon receipt of any $\text{LTC}_j$, the RTI reevaluates whether to send a TAG$_i$ or $\text{PTAG}_i$ to any downstream node $i$.
Next, whenever the RTI chooses to send a $\text{PTAG}_j$ to any node $j$ that is upstream of $i$, it should check if there is a pending $\text{NET}_i$ that it has not responded to.
If $G(\text{PTAG}_j) \ge g(\text{NET}_i)$, then it may be possible to send a TAG$_i$ or $\text{PTAG}_i$.
Specifically, if for every upstream $j$, the RTI has either a received $\text{LTC}_j$ with $g(\text{LTC}_j) \ge \text{NET}_i$
or has sent a $\text{PTAG}_j$ with $g(\text{PTAG}_j) \ge \text{NET}_i$ or TAG$_j$ with $g(\text{TAG}_j) \ge \text{NET}_i$,
then it can send either TAG$_i$ or $\text{PTAG}_i$.
It must send $\text{PTAG}_i$ if (and only if) at least one of the upstream $j$ is one for which the RTI has sent $\text{PTAG}_j$ with $G(\text{PTAG}_j) = g(\text{NET}_i)$. Otherwise, it can send TAG$_i$.

Although the bookkeeping may seem complicated, it is actually simple to implement.
Upon receiving updates from nodes ($\text{LTC}_j$ or $\text{NET}_j$) and upon making a decision to grant a TAG$_j$ or $\text{PTAG}_j$, the RTI needs to evaluate a simple expression for each downstream node $i$ if that downstream node has a pending $\text{NET}_i$ to which the RTI has not already responded.

\subsection{Inferring Port Status From Topology}
There are two situations that can lead to $B_i = G(\text{NET}_i)$. One reason is a coincidence that some node $j$ upstream of $i$ has sent to the RTI a NET$_j$ with $G(\text{NET}_j) = G(\text{NET}_i)$ and has not yet completed processing of that event.
The second reason is that $i$ is upstream of itself with a ZDC.
In both cases, it is valid for the RTI to issue a PTAG, but in the first case, it is not necessary.
In the second case, it \emph{is} necessary because without the PTAG, the program will be deadlocked.
\fixme{Give examples here.}

PTAG handling is considerably more complex than TAG handling.  It requires that nodes send out ABS messages when they have determined at a tag that no output will be produced, which increases network traffic and requires additional processing at the receiving end.
For this reason, we prefer to issue PTAG signals only when there is a ZDC.

Whether there is a cycle with no \code{after} delays, i.e.,~a path from a node $i$ back to itself with no \code{after} delays, can be determined while calculating $\mathcal{D}_{ij}$ for pairs of nodes $i$ and $j$ using Dijkstra's algorithm.
Each node $i$ can be marked in the RTI with a flag that can then be used to determine whether to issue a PTAG or wait until a TAG can be issued.

\subsection{In-Transit Message Queue}\label{sec:in-transit}

\begin{figure}
	\centering
	\includegraphics[width=0.95\columnwidth]{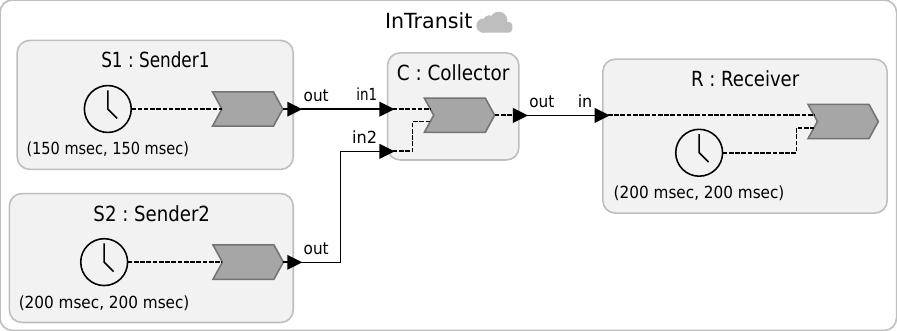}
	\caption{An example LF program where an RTI without $Q_i$ can give a wrong TAG$_{R}$ to the node $R$.}
	\label{fig:InTransit}
\end{figure}

In this section, we will prove by contradiction that the in-transit message queue $Q_i$ is necessary for each node $i$.
Assume that a new RTI, RTI$^\prime$ does not maintain $Q_i$ to forecast the earliest event of node $i$.
In this scenario, $E_j^\prime = \min(N_j, B_j^\prime)$ and $B_i^\prime = \min_{j \in U_i} D( \min(N_j, B_j^\prime), D_{ij})$ where $U_i$ is the set of $i$'s upstream nodes.
We will show that the RTI$^\prime$ mispredicts the future behavior of nodes of the program in \figurename~\ref{fig:InTransit} due to the absence of the in-transit message queues. 

\begin{figure}
\centering
\includegraphics[width=0.7\columnwidth]{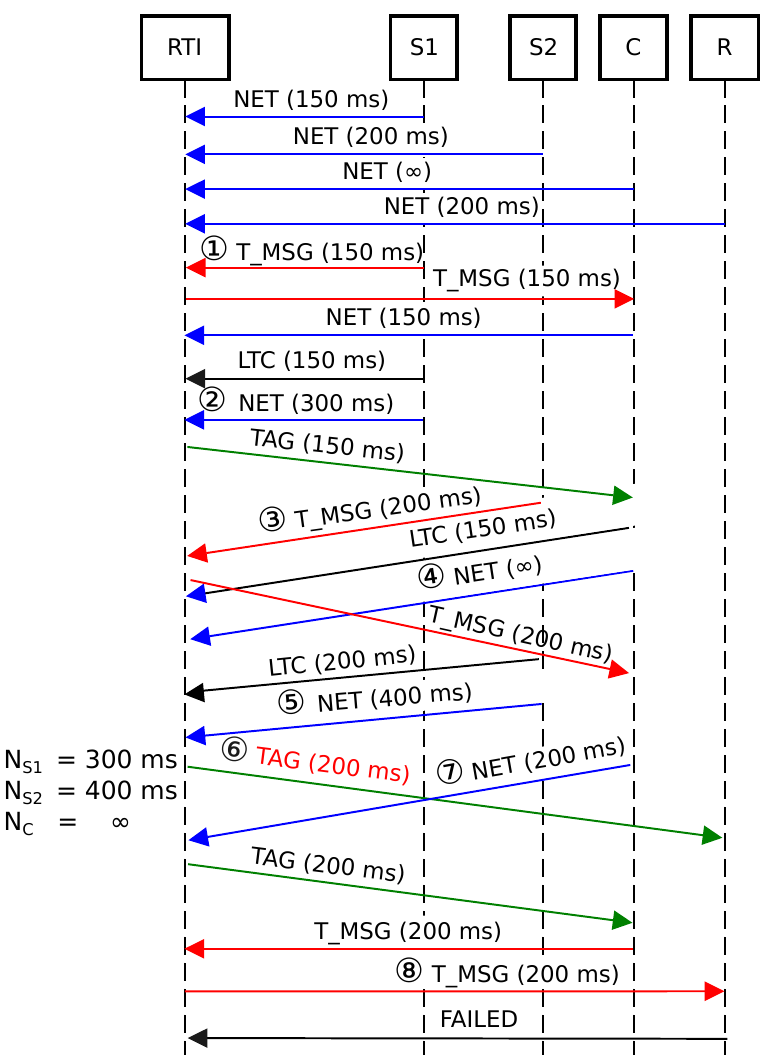}
\caption{A possible execution trace of the LF program in \figurename~\ref{fig:InTransit}.}
\label{fig:InTransit_trace}
\end{figure}

\figurename~\ref{fig:InTransit_trace} illustrates one of the possible execution
traces of the program in \figurename~\ref{fig:InTransit}. \code{S1}, \code{S2}, \code{C}, and \code{R} denote instances of \code{Sender1},
\code{Sender2}, \code{Collector}, and \code{Receiver}, respectively. In this trace, \code{R}
receives the message \textcircled{8} with a tag $G(\text{MSG}_{RC}) = 200~ms$ after it
has received the TAG$_R$ signal \textcircled{6} with a tag $G(\text{TAG}_R) = 200~ms$,
which leads to an incorrect behavior of \code{R}. (Here, we assume all microsteps are
0, so we do not show them.) A TAG$_R$ signal with a tag $G(\text{TAG}_R) = 200~ms$ is a
promise by the RTI that \code{R} has seen all messages with tags 200~ms or less,
which in this case is not true.

Let us follow the trace step by step to see what causes the RTI to send the wrong TAG$_R$.
At the start of the execution, $N_C$ is $\infty$.
After the RTI receives \textcircled{1}, it updates $N_C$ to $150~ms$ as $\min(150~ms, \infty)$ is $150~ms$.
However, it does not update $N_C$ to $200~ms$ when it sees \textcircled{3} because $200~ms$ is later than $150~ms$.
Then, $N_C$ is renewed to $\infty$ when the RTI sees \textcircled{4}. 
Note that \textcircled{3} and \textcircled{4} can arrive at the RTI in a nondeterministic order.
In this trace, \textcircled{3} arrives before \textcircled{4}.
After the RTI receives the signal \textcircled{5}, $N_{S1}$ and $N_{S2}$ are $300~ms$ and $400~ms$ because of \textcircled{2} and \textcircled{5}.

At this time, $B_R^\prime$ is $\min_{j \in {S1, S2, C}} D(\min(N_j, B_j^\prime), D_{Rj})$ = $\min_{j \in {S1, S2, C}} \min(N_j, B_j^\prime)$ ($\because \forall D_{Rj} = (0, 0)$).
According to the trace, $N_{S1}$ is $300~ms$, $N_{S2}$ is $400~ms$, and $N_C$ is FOREVER.
Also, $B_{S1}^\prime$ and $B_{S2}^\prime$ are {FOREVER} and $B_C^\prime$ is $300~ms$ (= $\min(300~ms, 400~ms)$).
As a consequence, the RTI$^\prime$ calculates $B_{R}^\prime$ as $300~ms$ and incorrectly grants TAG$_R$ with $200~ms$.
Due to the wrong TAG$_R$, \code{R} may proceed to $200~ms$ before \code{R} receives $\text{MSG}_{RC}$, thus \code{R}'s behavior becomes nondeterministic depending on whether \code{R} receives $\text{MSG}_{RC}$ before it proceeds to $200~ms$ or not. 
With $Q_i$, the RTI can recognize that \code{C} has an event at $200~ms$ (i.e., $B_C$ is $200~ms$) and successfully prevent this error.

\section{Provisional Tag Advance Grants}\label{sec:blocking}

This section discusses how we handle the PTAG and the ordering of reactions within each node as well as across nodes in a federated LF program.
We begin with two constraints on the execution order of reactions when handling messages~\cite{LohstrohEtAl:21:Towards}.
If multiple reactions are triggered at the same logical time:
\begin{enumerate}
	\item reactions in the same reactor must be executed in the order in which they are defined. \label{reaction-order1}
	\item if there is a zero-delay connection from a reaction ${r1}$ of one reactor to a reaction $r2$ of another reactor, $r1$ must complete its execution before $r2$ is triggered.\label{reaction-order2}
\end{enumerate}

In LF, an acyclic precedence graph called the reaction graph is generated to capture the constraints of the order of execution of reactions.
To avoid traversing the graph at runtime, we use a topological sort algorithm and assign levels to each reaction based on the topological depth~\cite{Menard:23:High-performance}.
This level is the same as the top level defined in~\cite{Kwok:99:Level} and provides a simple scheduling policy where a reaction is triggered only after reactions with lower levels have been completed.

\begin{figure}
	\centering
	\includegraphics[width=0.5\columnwidth]{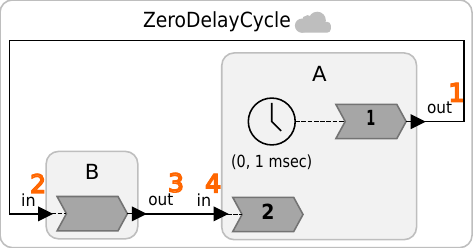}
	\caption{A simple program with a ZDC. TPO is marked orange.}
	\label{fig:ZDC}
\end{figure}

When a node is in a ZDC, there are two challenges to meeting the rules for executing reactions.
First, a node must wait for network input ports' statuses to become known before executing a reaction that depends on the input ports.
Let us consider a simple federated LF program with a ZDC depicted in Figure~\ref{fig:ZDC}.
After starting the program, $A$ will receive $\text{PTAG}_A$ where $G(\text{PTAG}_A)$ is equal to the start tag. It will then execute reaction 1, which does not depend on the input port, allowing it to produce an output.
Next, $A$ needs to block its execution until the input port's status becomes known.
The status becomes known after $B$ sends either $\text{MSG}_{AB}$ or $\text{ABS}_{AB}$.
Section~\ref{sec:MLAA} explains how we tackle this challenge efficiently using a new concept that we call Max Level Allowed to Advance (MLAA).

Second, preserving the dependency between reactions from different nodes is challenging.
In LF, each node is decomposed into a single program and compiled individually.
Thus, we will lose some dependencies between nodes.
For example, in Figure~\ref{fig:ZDC}, once the two programs have been separated, $A$ does not know there is a dependency between $A$ and $B$.
In Section~\ref{sec:TPO}, we will show that the lost information can lead to a significant problem (deadlock) and will introduce a Total Port Ordering (TPO) level to solve the problem.

\subsection{Max Level Allowed to Advance}\label{sec:MLAA}

\begin{figure}
	\centering
	\includegraphics[width=0.9\columnwidth]{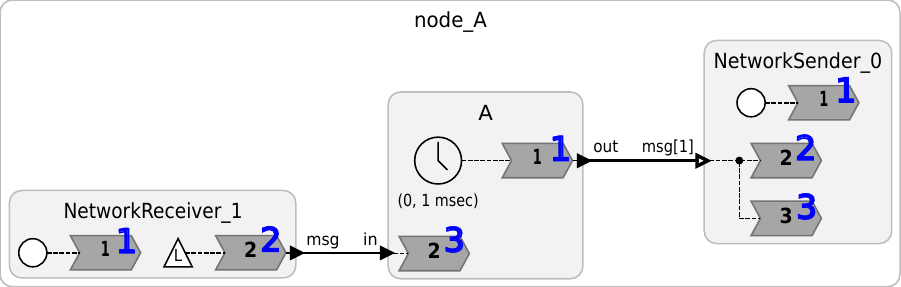}
	\caption{A separate program for the node $A$ in Figure~\ref{fig:ZDC}. The levels of reactions are marked blue.}
	\label{fig:NR}
\end{figure}

When a federation is decomposed into separate nodes, special-purposed reactors called network receivers and network senders are generated for each network input and output port.
The diagram in \figurename~\ref{fig:NR} shows the network receiver and sender generated for the node $A$ in \figurename~\ref{fig:ZDC}.
When a $\text{MSG}_{AB}$ arrives at input port \code{A.in}, the network input reaction of the port (reaction 2 of the network receiver) is triggered.
The reaction that depends on \code{A.in} (reaction 2 of $A$) is triggered by the network input reaction.
The network sender sends $\text{MSG}_{BA}$ to the RTI using its reaction 2 if reaction 1 of $A$ produces an output or sends $\text{ABS}_{BA}$ using its reaction 3 if the reaction does not produce an output.

At each tag, a node executes triggered reactions in order of their level, possibly executing reactions in parallel on multiple cores when they have the same level.
When there is a reaction $r1$ that depends on an input port for which the status is unknown, the node must block before executing the reaction until the port status becomes known.
Additionally, every reaction with a level higher than $r1$ must also be blocked because if $r1$ is triggered after executing those reactions, it will have violated the level order execution.
Note that a port can be unknown only if its connection is a part of a ZDC. This is because if there weren't a ZDC, we would have simply waited for the port's upstream federate to advance to a sufficiently late tag to make the port known at the current tag.

To do the blocking, each node calculates an MLAA to know what reactions can be executed.
The MLAA is the minimum level of an unknown port's network input reaction, and it acts like a barrier that prevents advancing to the next level until the port's status becomes known.
At runtime, a node can execute a reaction only if its level is less than the MLAA.
Assume that there is a node $i$ that has $n$ network input ports that are part of a ZDC, and that the levels of the dedicated network input reactions for the ports are $l_1$, $l_2$, ..., $l_n$.
Also assume that every port status is unknown when advancing the current tag to tag $g$.
Then node $i$ will initialize the MLAA to $\min\limits_{\forall k \in \{1..n\}} l_k$ when it advances its tag to $g$; this is the lowest level of the network input reactions.
When the status of $l_m$'s port becomes known, the node recalculates MLAA, this time ignoring $l_m$, i.e., $\min\limits_{\forall k \in \{1..n\} \setminus \{m\}} l_k$.
If every port becomes known, MLAA is set to $\infty$ so that the node can execute every triggered reactions and advance to the next tag.

Note that this strategy is approximate and can result in some unnecessary blocking of reactions that do not depend on unknown input ports just because their level equals or exceeds the level of a blocked network input reaction.
A less restrictive solution could be constructed by using, at runtime, a full representation of the reaction graph rather than its abstraction using levels.
We show here, however, that this elaboration is not necessary to achieve deadlock-free execution of any network program that is free of causality loops.
The additional complexity could improve performance, but TPO ensures that it will not affect correctness.

\subsection{Total Port Ordering}\label{sec:TPO}

The level-based abstraction of the reaction graph is an over-approximation in the sense that it implies an ordering relation in the execution of reactions within a given tag that is a superset of the ordering relation implied by the reaction graph. 
This is why, in the presence of individually compiled nodes, it is non-obvious that the independently computed level assignments for the different nodes will be composed in such a way that the resulting global ordering relation is cycle-free or, equivalently, deadlock-free.


We will show the program can be deadlocked if the nodes assign levels independently (without respecting to globally-defined dependency).
The blue numbers in \figurename~\ref{fig:NR} illustrate the levels assigned for reactions.
The minimum level is 1.
The level of reaction 2 in the network receiver is 2 because reaction 1 is defined before the network input reaction.
Reaction 2 in $A$'s level is 3 because it depends on the level 2 reaction.

At the start tag, the level of unknown port \code{A.in}'s network input reaction is 2, so the node initializes MLAA to 2.
Thus, the node blocks its execution after processing level 1 reactions and waits until port \code{A.in}'s status becomes known.
However, \code{A.in}'s status cannot become known, and the program deadlocks.
\code{A.in} depends on the reaction in $B$ in \figurename~\ref{fig:ZDC}, and the node $B$ can execute its reaction after it receives $\text{MSG}_{BA}$ or $\text{ABS}_{BA}$ from reaction 2 or reaction 3 in the network sender of $A$.
The reactions are also blocked by MLAA, and MLAA is updated if $A$ receives network input from $B$.
This means that $A$ and $B$ are waiting for each other, resulting in a deadlock.

Total port ordering (TPO) solves this by fully specifying a globally agreed-upon ordering that must be respected at the boundaries between nodes.
TPO is computed before the individual nodes are compiled and then respected by each individual node so that deadlock freedom can be ensured without any one federate requiring global knowledge of the other nodes and their internals.

To compute TPO, we first compute the ordering that comes from a global level-based topological sort of the reactions on the basis of the reaction graph. 
Then, we define the TPO to be the subset of that ordering that only relates the ports of federates with each other.
The orange numbers in \figurename~\ref{fig:ZDC} denote the computed TPO for that example.


\begin{figure}
	\centering
	\includegraphics[width=0.9\columnwidth]{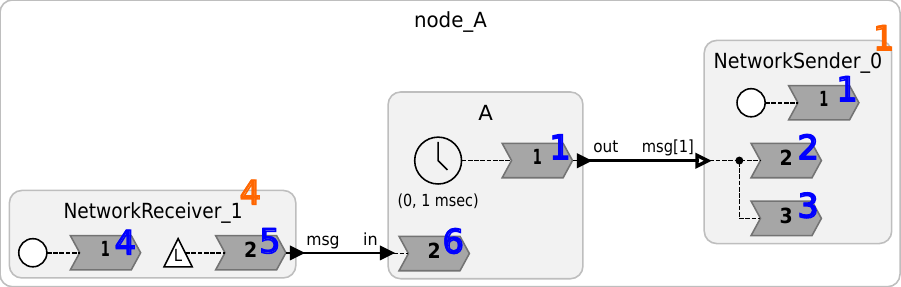}
	\caption{A separate program for the node $A$ in Figure~\ref{fig:ZDC}. Levels (blue) are assigned to reactions with respect to TPO (orange).}
	\label{fig:LevelbyTPO}
\end{figure}
For each node, there exists at least one valid level assignment (namely, the level assignment that the node had in the global level-based sort) that is consistent with the TPO; this guarantees that when it is separately compiled, that node will be able to find some level assignment for itself that is also valid (although it may be slightly different).
The restriction of the TPO to a node's ports fully characterizes the node's externally observable blocking behavior; this guarantees that, since the global sort was cycle-free (equivalently, deadlock-free), the composition of the nodes with their individually re-computed level assignments will be deadlock-free as well.
This aligns with the statement mentioned in Section~\ref{sec:MLAA} that using the level-based abstraction of dependency graph with total port ordering does not affect correctness.

\figurename~\ref{fig:LevelbyTPO} shows how $A$ computes levels of reactions with respect to assigned TPO.
With TPO, $A$ knows that the input port \code{A.in} depends on the output port \code{A.out}.
Thus, the levels of the reactions related to \code{A.in} becomes higher than the reactions related to \code{A.out}, resulting in resolving the deadlock by allowing the network sender to send $\text{MSG}_{BA}$ or $\text{ABS}_{BA}$.

\section{Evaluation}\label{sec:evaluation}




In this section, we estimate the possible performance impact of our proposed approach to handling ZDCs.
Specifically, we evaluate our approach in comparison with the conventional approach of inserting microstep delays using LF.
As a benchmark, we take the simple federated example with a ZDC shown in \figurename~\ref{fig:ZDC} and the almost same example with a microstep delay put on the feedback path from \code{A.out} to \code{B.in}.

We run experiments under two types of network conditions.
In one setup, we connect the RTI and federates over the same local area network (LAN).
In another setup, we connect the federates to the RTI via a Wi-Fi gateway. 
In the LAN setup, the network round trip time measured as a network ping is 0.483 ms on average, while the round trip time measured for the Wi-Fi is 8.681 ms on average.
We note that the actual network round trip time varied over time, especially with the Wi-Fi setup, due to the dynamic network environments, which we can also see in real-world wireless network environments.

\begin{figure}
	\centering
	\includegraphics[width=0.95\columnwidth]{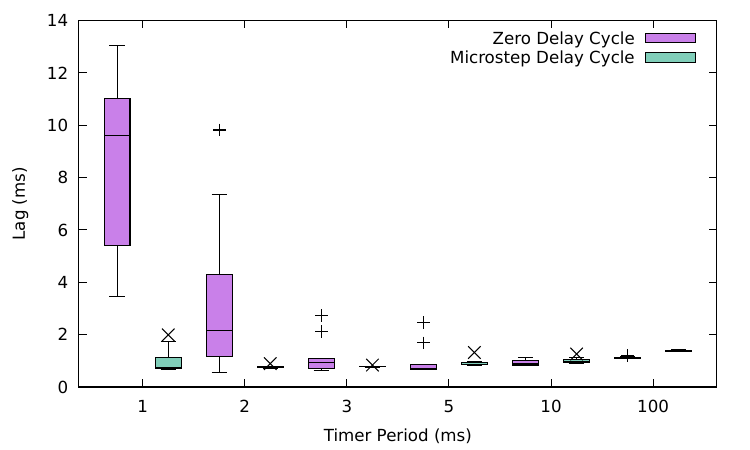}
	\caption{Lags measured under the LAN setup with an average network round trip time of 0.417 ms.}
	\label{fig:LagCable}
\end{figure}

\begin{figure}
	\centering
	\includegraphics[width=0.95\columnwidth]{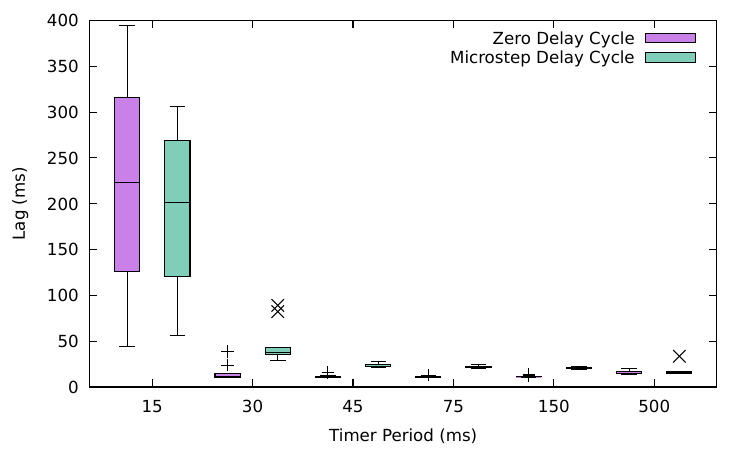}
	\caption{Lags measured under the Wi-Fi setup with an average network round trip time of 8.681 ms.}
	\label{fig:LagWiFi}
\end{figure}

For the two network conditions, LAN and Wi-Fi, we measure the lag, which we define as the time difference between the physical time and the logical time of the final reaction of the cycle, reaction 2 of Reactor A.
We run the experiments while varying the period of the timer in Reactor A.
While running the experiments with shorter timer periods, we noticed the lag starts getting accumulated to the point that the federated program cannot guarantee the intended timing behavior.
In this section, we call the timer period when the program starts to break down in terms of timing behavior, a \textit{breakdown period}.



\figurename~\ref{fig:LagCable} and \figurename~\ref{fig:LagWiFi} illustrate the measured lags of the Lingua Franca programs in \figurename~\ref{fig:ZDC} (our approach, marked purple) and the same program with a microstep delay on the feedback path from \code{A.out} to \code{B.in} (marked green)
under LAN setup and Wi-Fi setup, respectively.
For each timer period (on the x-axis), we set the duration of the experiment so that Reactor A sends  500 messages in total. For example, when the timer period is 1 msec, we run the experiment for 500 msec; for the timer period of 2 msec, the experiment runs for 1,000 msec, and so on.
We divided the duration of each experiment into 10 intervals and measured the average lag for each interval, as we observed that the lag can change over time, even in each run of experiments.
We also ran 10 runs for the same experiment and averaged the lags for each interval.

In the LAN setup shown in \figurename~\ref{fig:LagCable}, the breakdown period is 2 ms for the proposed approach and 1 ms for the microstep-based approach. 
In the Wi-Fi setup shown in  \figurename~\ref{fig:LagCable}, the breakdown period is around 15 ms for both approaches.
In terms of the lag, our approach had a similar breakdown period, and until we reached the breakdown period, both approaches showed similar average lags.
Thus,  we can conclude that the proposed zero delay cycle handling does not incur any significant overhead in terms of the lag of the reactions.




\section{Conclusions}\label{sec:conclusions}

Most distributed discrete-event frameworks prohibit zero-delay cycles.
We show that this prohibition is not necessary and give example program patterns where it is more convenient
for the programmer to specify the desired functionality with ZDCs than to work around the prohibition.
We derived the information needed to coordinate the execution of distributed components in zero-delay cycles
and give an implementation that extends the open-source Lingua Franca coordination language.
Our experimental evaluation suggests that the performance cost of removing this prohibition is modest.

\fixme{acks}



%

\bibliographystyle{IEEEtran}
\bibliography{Refs}

\vspace{11pt}


\vfill

\end{document}